\documentclass[prl,twocolumn,superscriptaddress,showpacs,floatfix,amsfonts]{revtex4}
\usepackage{graphicx,graphics,color,epsfig,exscale}
\usepackage{bm}
\usepackage{amsmath}
\usepackage{amssymb}
 
\def\Im{\text{Im}} 
\def\sgn{\text{sgn}}

\def\be{\begin{equation}} 
\def\ee{\end{equation}} 
\def\bea{\begin{eqnarray}} 
\def\eea{\end{eqnarray}}
\begin{document}
\preprint{}
\title{Quantum Criticality out of Equilibrium: 
Steady State in a Magnetic Single-Electron Transistor}

\author{Stefan Kirchner}
\affiliation{Department of Physics \& Astronomy, Rice University,
Houston, Texas, 77005, USA}
\affiliation{Max Planck Institute for the Physics of Complex Systems,
01187 Dresden, Germany}
\author{Qimiao Si}
\affiliation{Department of Physics \& Astronomy, Rice University,
Houston, Texas, 77005, USA}

\begin{abstract}
Quantum critical systems out of equilibrium are of extensive
interest, but are difficult to study theoretically. 
We consider here the
steady state limit of a single-electron transistor
with ferromagnetic leads.
In equilibrium ({\it i.e.}, bias voltage $V=0$),
this system 
features a continuous quantum phase transition
with a critical destruction of the Kondo effect.
We construct 
an exact quantum Boltzmann treatment in a dynamical large-N limit,
and 
determine the universal scaling functions 
of 
both the 
non-linear conductance
and fluctuation-dissipation ratios.
We also elucidate 
the 
decoherence properties
as encoded in the local
spin response.
\end{abstract}
\maketitle

Quantum phase transitions are of extensive current interest in 
a variety of strongly correlated electronic and atomic
systems\cite{Natphysi_focus}. A quantum critical point
occurs when such transitions are second order\cite{Sachdev}.
In a quantum critical state, 
there is no 
intrinsic energy scale in the excitation spectrum.
External probes will readily drive the system out of equilibrium,
such that the standard notion
of linear response
breaks down.
The response of such a system to an external
drive cannot be directly related 
to its intrinsic fluctuations in equilibrium.

Compared to
their classical counterparts, quantum critical systems are 
more
difficult to study already at the equilibrium level.
A quantum critical point involves the mixing
of statics and dynamics. This complicates the determination of the 
equilibrium fluctuation spectrum, especially in the long-time 
``quantum relaxational'' regime ($\hbar \omega \ll k_B T$) 
at finite temperatures\cite{Sachdev}.
In addition, the classification
of the universality classes
is
delicate, since the critical modes may incorporate
inherently quantum ones that go beyond the order-parameter
fluctuations. A typical example is for Kondo systems, in which quantum
criticality can involve a critical destruction of the Kondo 
entanglement\cite{Si.01}. 

To make progress theoretically, we need approaches
that can, on the one hand, capture the scaling properties 
of a quantum critical point in equilibrium, and, on the other hand,
be
extended to settings out of equilibrium.
This is impeded by the lack of a general understanding of how to generalize the free energy functional to
stationary nonequilibrium states from which scaling relations could be obtained.
Some progress has been made in recent years.
In the case of the two-dimensional superconductor-insulator transition,
the steady-state conductance in the non-linear regime
[$V (\gg k_B T /e) \rightarrow 0$]
is found to differ from its counterpart
in the linear-response regime 
[$T (\gg e V /k_B) \rightarrow 0$]~\cite{Green.05,Dalidovich.04},
in analogy to the non-commutativity of the 
$\omega \rightarrow 0$ and $T \rightarrow 0$
limits of the equilibrium current-current correlation
function\cite{Damle.97}.
A complementary issue, {\it viz.} quantum criticality induced by
a steady-state voltage drive,
has also been addressed in 
the contexts of itinerant 
magnetism\cite{Mitra.06,Feldman.05}.
Still, there is a considerable need for 
insights into many open issues regarding quantum critical
systems out of equilibrium.
How do we connect the scaling behavior of the equilibrium fluctuation
spectrum with that of the out-of-equilibrium properties?
How to calculate the
universal out-of-equilibrium scaling functions?
Is the fluctuation-dissipation theorem violated in a universal
way in the quantum critical regime?

In this letter, we study 
these issues in the steady state of a quantum critical system 
out of equilibrium.
We have chosen a ferromagnetic single-electron 
transistor (SET)\cite{Pasupathy.04} as a case study.
It was shown earlier that this system undergoes a 
gate-voltage-induced quantum
phase transition,
which is characterized by a continuous suppression of 
the Kondo screening\cite{Kirchner.05a,Kirchner.08b}.
A dynamical large-N limit was found to capture the universal properties
of the quantum criticality in equilibrium\cite{Kirchner.05a,Zhu.04}.
Here, we show that this approach can be reliably generalized to 
the case of steady states under a finite bias, and use it to determine
the full scaling functions
of non-equilibrium properties.
Our results establish the system
as a prototype model setting 
in which the general issues raised earlier can be systematically
studied.
In addition to elucidating the nonequilibrium quantum criticality,
our work also contributes to the general understanding of the Kondo effect
out of 
equilibrium\cite{Wingreen.94,Kaminski.00,Coleman.01a,Rosch.01,Doyon.06,Utsumi.05}.

{\it The system, the model, and its large-N generalization:~}
The system we consider is illustrated in Fig.~1a, showing a quantum dot
coupled to two leads of ferromagnetic metals whose magnetizations are 
antiparallel.
It was demonstrated experimentally\cite{Pasupathy.04} and theoretically 
\cite{Martinek.03,Choi.04} 
that a Kondo effect can
still occur in the Coulomb-blockade regime,
in spite of a large Stoner splitting
of the conduction electron bands.
It was shown\cite{Kirchner.05a,Kirchner.08b}
that the system should be modeled
in terms of a Bose-Fermi Kondo Hamiltonian.
A localized 
moment on the dot 
is coupled to a fermionic bath, corresponding to the Stoner continuum
of the ferromagnetic leads, with a coupling constant $J$.
Simultaneously, it is coupled to a bosonic bath,
associated with the magnons of the 
leads,
with a coupling constant $g \approx J^2/\Gamma$ (where 
$\Gamma$ is the bare resonance width). Tuning the gate voltage varies
the ratio ($g/J$) of the two couplings, giving rise to 
a continuous transition between the Kondo phase and 
a critical local moment (LM) phase. 

Of interest to us here
is the universal behavior of the quantum critical point, 
referred to henceforth 
as C, and the quantum critical LM phase.
The equilibrium behavior is illustrated in Fig.~1b, for a fixed
$J$ (and, equivalently, a fixed bare Kondo scale
$T_K^0$);
the quantum critical 
point C corresponds to $g=g_c$, and the quantum critical LM phase
occurs at $g>g_c$.
Universal scaling behavior in this model sets in whenever the probing
energy and the temperature are smaller than $\min[T_K^0,\Lambda_s]$,
where
$\Lambda_s$ is
the energy at which the magnon dispersion merges with the
Stoner continuum. Physically, we have
$T_K^0 \ll \Lambda_s$ and we therefore restrict
ourselves to  $T \ll T_K^0$.

Anticipating the large-N limit to be taken,
we write the effective low-energy model 
in the following form:
\begin{eqnarray}
H&=&\sum_{i, \sigma, \sigma'} (J/N) 
{S}^\alpha \cdot 
c^\dagger_{i \sigma } { \tau^\alpha}_{\sigma \sigma'} c^{}_{i \sigma'} + 
H_0(c)
\nonumber \\
&+& (g/\sqrt{N}) 
{\bm S}
\cdot {\bm \Phi}
+\sum_q \omega_q {\bm \Phi}_q^{\dagger}\cdot{\bm \Phi}_q^{} .
\label{eq:MC-Hamilton}
\end{eqnarray}
Here, $c^\dagger_{i,\sigma} 
= (t_L^{} c^\dagger_{i \sigma,L } +
t_R^{} c^\dagger_{i \sigma,R })/\sqrt{|t_L|^2+|t_R|^2}$
creates an electron in 
a linear combination of the local states in the leads for spin $\sigma$
and channel index $i$~;
$\sigma,\sigma'=1,\ldots,N$,
$i=1,\ldots,M$,
and $t_L$ and $t_R$ are the 
hybridization matrix elements with the left and right
leads, respectively.
$H_0(c) = \sum_{\alpha=L,R}H_{0\alpha}(c_{\alpha})$,
with 
$H_{0\alpha}(c_{\alpha}) = \sum_{k i \sigma} 
(\epsilon_{k,\sigma,\alpha}-\mu_{\alpha}) 
c_{{\bm k} i \sigma,\alpha}^{\dagger} c_{{\bm k} i \sigma,\alpha}^{}$.
The energy dispersion, $\epsilon_{k,\sigma,\alpha}$,
contains the appropriate Zeeman shifting,
but is otherwise taken to be featureless and has in particular a finite
density of states in between the two chemical potentials $\mu_L$ and $\mu_R$.
${\bm \Phi}_{}^{}= \sum_q ({\bm \Phi}_q^{\dagger}+{\bm \Phi}_q^{})$ 
is also a linear combination of the spin waves on the left and right
leads; it has been written to contain $N^2-1$ components. 
The matrices ${\tau^\alpha}, \alpha=1,\ldots,N^2-1$ span the 
generators of SU(N).
For simplicity, we consider $t_L=t_R$
(otherwise, a local magnetic field will
arise\cite{Martinek.03,Kirchner.08b}).
It follows from $\omega_q \propto q^2$
that 
the spectrum of the bosonic modes is
\begin{equation}
\sum_q \left [
\delta(\omega-\omega_q)-\delta(\omega+\omega_q)
\right ] \sim |\omega|^{1-\epsilon}
sgn(\omega),
\label{eq:subohmic}
\end{equation}
with $\epsilon=1/2$.
(Our results can be easily generalized
to the generic sub-Ohmic case, $0<\epsilon<1$.)
The power-law form given
in Eq.~(\ref{eq:subohmic})
is valid
for $\omega$ below some cutoff
frequency, 
$\Lambda$.
For simplicity, we 
consider the particle-hole symmetric case without any
potential-scattering term.
The latter is 
exactly marginal\cite{Zhu.02,Zarand.02,Kirchner.05a},
and 
would hence only add a constant offset
without modifying the $T$- and $V$-dependent scaling terms.
The explicit Keldysh calculation in the presence of the 
potential scattering term is more difficult, because  
solving the pseudofermion constraint in the absence of particle-hole symmetry
is more involved;
we have nonetheless used such a calculation to confirm 
the above\cite{inpreparation}.
\begin{figure}[t!]
\includegraphics[width=0.5\textwidth]{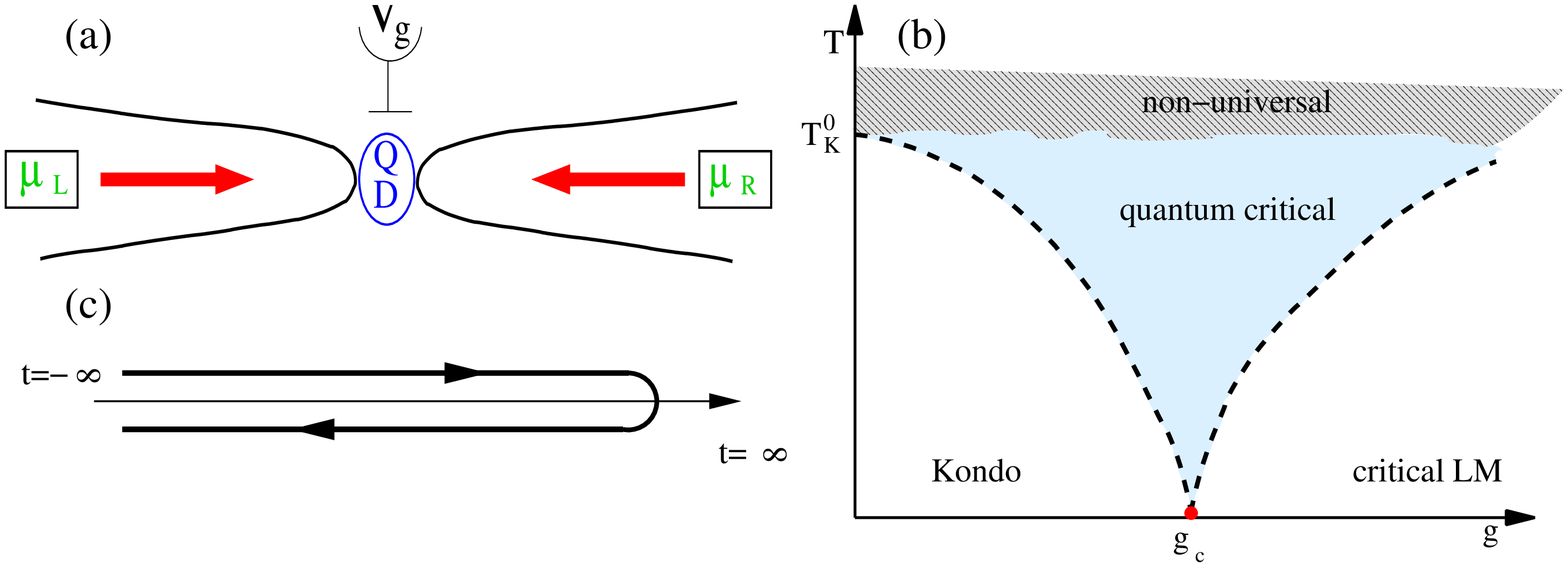}
\caption{(a) A sketch of the magnetic SET.
(b) The phase diagram of the magnetic transistor in equilibrium
($\mu_R=\mu_L$).
$T_K^0$ is the bare Kondo scale, appearing when $g=0$,
{\it i.e.}, when there is only the coupling $J$.
A critical point at coupling $g_c$ separates the Kondo phase
from a critical local moment (LM) phase.
(c) 
The Keldysh contour appropriate for the non-equilibrium cases.
}
\label{FIG1}
\end{figure}

The dynamical large-N limit
in equilibrium
was considered in \cite{Zhu.04},
and this limit has been shown to correctly capture the 
dynamical scaling properties (including, in particular,
the finite-temperature relaxational properties)
of the $N=2$ case for both the 
quantum critical point ($g=g_c$) and 
quantum critical LM phase ($g>g_c$)\cite{Kirchner.05a,Zhu.04}.
In the large-N limit\cite{Parcollet.98,Cox.93},
${\bm S}$ is expressed in terms of pseudofermions, 
$S_{\sigma \sigma'}=f^{\dagger}_{\sigma}
f^{}_{\sigma'}-\delta_{\sigma \sigma'}q_0^{}$, 
where $q_0=Q/N$ is related to
the chosen irreducible representation of SU(N) and it also 
appears in the
pseudo-fermion occupancy 
constraint $\sum_{\sigma}f^{\dagger}_{\sigma}f^{}_{\sigma}=Q$.
The quartic interaction 
is decoupled 
via a Hubbard-Stratonovich field $B_{i}$ ($i=1,\ldots,M$),
which is conjugate to
$\sum_{\sigma}c^{\dagger}_{i \sigma}f^{}_{\sigma}/\sqrt{N}$.

For a finite bias voltage,
we work on the Keldysh contour, depicted in Fig.~\ref{FIG1}c.
Since we are interested in the steady state limit, we specify 
the state of the system at the infinite past $t_0=-\infty$, 
so that at finite time all initial correlations will have washed out.
A finite bias voltage, $eV=\mu_L-\mu_R$, is incorporated in the 
following  lesser
function for the fermionic bath\cite{Wingreen.94,Hettler.94}:
\begin{equation} 
\mathcal{G}_c^{<}(\omega)=
\pi i [ f_R(\omega) + f_L(\omega)] \, \rho(\omega),
\label{eq:distrib}
\end{equation}
with 
$f_{L/R}(x)=\left [\exp((x-\mu_{L/R})/T)+1 \right ]^{-1}$
and 
$\mu_L=eV/2=-\mu_R$. 
Eq.~(\ref{eq:distrib})
also defines
the lead distribution function.
Both leads are held at temperature $T$ which completely characterizes the magnon distribution function.
The magnons, therefore, retain thermal equilibrium, and 
the 
corresponding lesser function 
is
\begin{equation}
 \mathcal{G}_{\Phi}^{<}=-2\pi i b(\omega) A_{\Phi}(\omega) .
\end{equation}
Here, 
$A_{\Phi}(\omega) = [1/\Gamma(1/2) ]|\omega|^{1/2}\sgn(\omega)$, where 
$\Gamma$ is the Euler Gamma function,
and 
$b(\omega)= \left [ \exp(\omega/T)-1 \right ]^{-1}$.

{\it Large-N limit on the Keldysh contour:~}
In the equilibrium case,
the large-N limit yields saddle-point equations
for the pseudoparticle self-energies~\cite{Zhu.04}.
On the Keldysh contour,
the saddle point equations become 
$\Sigma_f(t) = -\kappa \,i\,\mathcal{G}_c(t) G_B(t)- 
g^2\, i\,\mathcal{G}_{\Phi}(t) G_f(t)$, and
$\Sigma_B(t) = i\, \mathcal{G}_c(t) G_f(-t)$,
with $\kappa \equiv M/N$. 
Performing 
an
analytical continuation onto the real time axis\cite{Haug,Kamenev.04},
we have
\begin{eqnarray}
\Sigma_f^{<}(t)\,&=&\, -\kappa \,i\,\mathcal{G}_c^{<}(t) G_B^{<}(t)\,-
\, g^2\,i\, \mathcal{G}_{\Phi}^{<}(t) G_f^{<}(t), \nonumber\\
\Sigma_f^{>}(t)\,&=&\,- \kappa \,i\,\mathcal{G}_c^{>}(t) G_B^{>}(t)\,-
\, g^2\, i\, \mathcal{G}_{\Phi}^{>}(t) G_f^{>}(t), \nonumber \\
\Sigma_B^{<}(t)\,&=&\, i\, \mathcal{G}_c^{<}(t) G_f^{>}(-t), \nonumber \\
\Sigma_B^{>}(t)\,&=&\, i\, \mathcal{G}_c^{>}(t) G_f^{<}(-t).
\end{eqnarray}
Retarded ($r$) and advanced ($a$) functions are related to the lesser and 
greater functions via
$G^{a}(t,t^\prime)=\Theta(t^\prime-t) (G^{<}(t^\prime-t) -G^{>}(t^\prime-t))$ and 
$G^{r}(t,t^\prime)=\Theta(t- t^\prime) (G^{>}(t^\prime-t) -G^{<}(t^\prime-t))$,
so that $G^{r}-G^{a}=-2 i \mbox{Im}(G)=G^{>}-G^{<}$.
\begin{figure}[t!]
\includegraphics[width=0.5\textwidth]{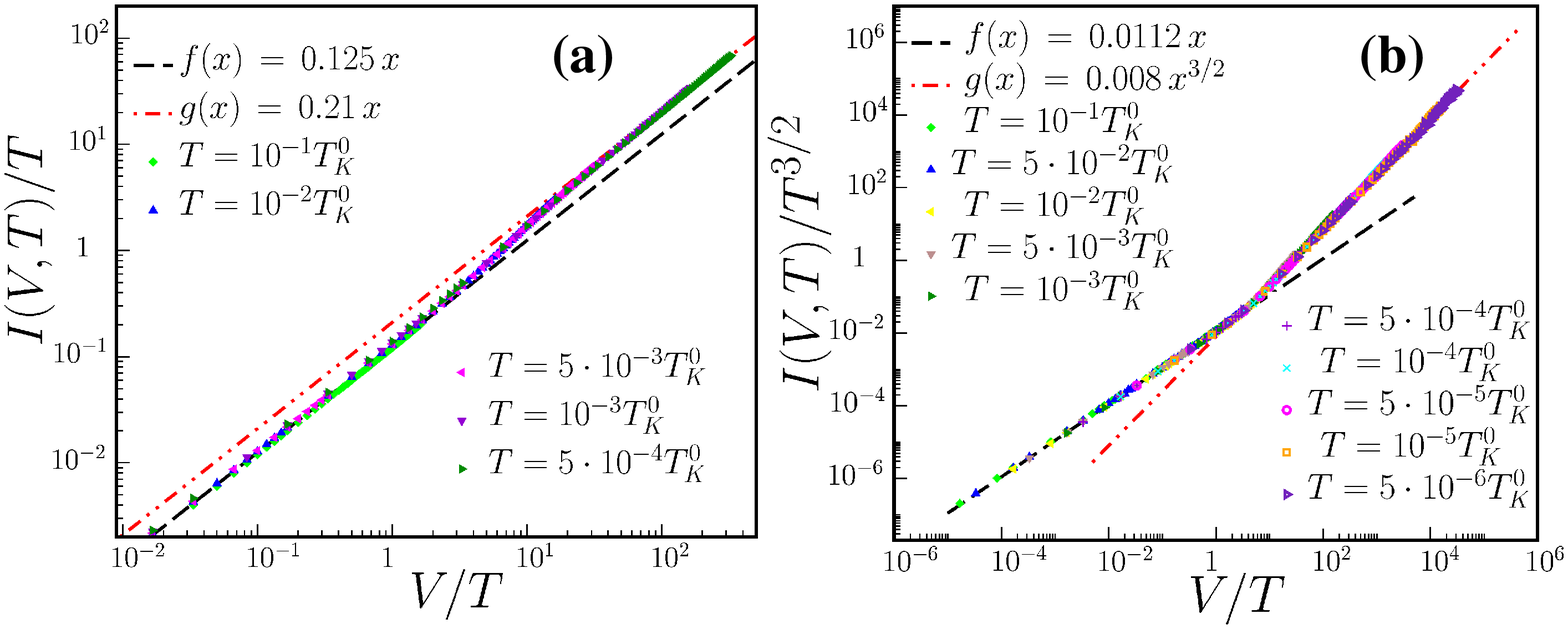}
\caption{Scaling of the current-voltage ('IV') characteristics
near (a) the critical (C) fixed point ($g=g_c$) 
and (b) the critical LM phase ($g=5g_c>g_c$).
In (a), we have
shown the results over a reduced range
of $V/T$ in order to highlight the crossover in the scaling function;
scaling is in fact 
seen
over a similarly extended range (nearly 10 decades)
of $V/T$ as shown in (b).
The results are shown with $\hbar$,$e$,$k_B$ being set to $1$.
}
\label{Figure2}
\end{figure}
$G^{<}\,=\, G^{r} \Sigma^{<} G^{a}$,
where $G^r$/$G^a$ 
are 
obtained from $G^{-1}(\omega)=G_0^{-1}(\omega)-\Sigma(\omega)$,
with 
$G^{-1}_{0,B}=1/J$ and
$G^{-1}_{0,f}=\omega-\lambda$;
$\lambda$ 
is a Lagrangian parameter enforcing 
the pseudo-fermion occupancy constraint.

We solve the saddle point equations numerically for given temperature $T$
and voltage $V$ on real frequencies. We choose $\kappa =1/2$ 
and $\rho_0(\omega)=\rho_0 \exp(-\omega^2/\pi)$.
The effective Kondo coupling is taken to be 
$J\rho_0=(J_{LL}+J_{RR})\rho_0=0.8/\pi$,
resulting in a bare nominal 
Kondo scale $T_K^0\rho_0=0.06/\pi$ at $g=0$.
The bosonic bath is cut off smoothly at $\Lambda\rho_0=0.05/\pi$.
This yields a critical coupling $g_c\rho_0 \approx 0.79/ \pi$.
For the LM regime we will illustrate the result at $g=5 g_c$,
which is sufficiently large to ensure an extensive scaling regime.
The convergence criterion is that,
for both
$G^{>}_{f}-G^{<}_{f}$ and   
$G^{>}_{B}-G^{<}_{B}$,
the sum over the entire frequency range of the relative difference
between two consecutive iterations is less than $10^{-5}$. 

{\it Steady-state current:~}
The current between the dot and the left lead for an arbitrary bias voltage 
$V$ is \cite{Meir.92} 
\begin{eqnarray*}
I_L&=&\frac{i e}{\hbar}\int d\omega \rho(\omega)
\big 
[f_L(\omega)({\mathcal{T}}^r_{LL}(\omega)-{\mathcal{T}}^a_{LL}(\omega))
+{\mathcal{T}}^<_{LL}(\omega) \big ]~. 
\label{eq:current}
\end{eqnarray*}
In the steady state limit, $I_L+I_R=0$,
and the 
current can then be alternatively cast into:
\begin{eqnarray}
I(V,T)&=&\frac{e}{2\,\hbar}\int d\omega\, \rho(\omega)
\left [f_L(\omega)-f_R(\omega) \right ] \nonumber \\
&\times& \mbox{Im}\big[ {\mathcal{T}}^{a}_{LL}(\omega,T,V)
+{\mathcal{T}}^{a}_{RR}(\omega,T,V)\big] .
\label{current_expression}
\end{eqnarray}
Here, ${\mathcal{T}}_{\alpha,\beta},~(\alpha,\beta=L/R)$,
is the T-matrix of the Bose-Fermi Kondo model.
In the large-N limit
we consider, ${\mathcal{T}}^{ }(t)\equiv {\mathcal{T}}^{ }_{LL}
+{\mathcal{T}}^{ }_{RR}=(i/N)\,G_B(-t)G_f(t)$.

{\it V/T scaling of the I-V characteristics:~}
We are now in a position to discuss the quantum critical behavior out 
of equilibrium.
Fig. \ref{Figure2}(a) shows the current-voltage characteristics 
at the quantum critical point, $g=g_c$.
In the linear-response regime, for $T\rightarrow 0$ after 
taking $V\rightarrow 0$,
the conductance is a universal constant in the leading
order\cite{Kirchner.05a}.
In the non-linear regime, for $V \rightarrow 0$ at $T=0$, 
we find that the conductance is also a universal constant,
but differs from the linear-response value. 
This is reminiscent of what happens\cite{Green.05} 
at the two-dimensional
superconductor-insulator
transition. It is to be contrasted with the result in the 
two-channel Kondo problem, in which the two 
constants
are identical; $V/T$ scaling is then analyzed in an
unorthodox fashion, in terms of 
$\Delta G \equiv G(V,T)-G(V=0,T=0)$\cite{Potok.07,Ralph.94,vDelft.99}.

Beyond these limits, we are able to determine
the conductance for arbitrary $V,T$. We find that 
$I/T$ shows a scaling collapse of $V/T$, provided $V,T < \sim T_K^0$.

Consider next the quantum critical LM phase at $g>g_c$. 
In the linear response regime, the conductance is 
proportional to $T^{1/2}$\cite{Kirchner.05a}.
In the non-linear regime, $V \rightarrow 0$ at $T=0$, we find that 
the conductance goes as $V^{1/2}$.
Fig. \ref{Figure2}(b) now demonstrates a scaling collapse
of $I/T^{3/2}$ vs. $V/T$. For both cases, Fig.~2 shows the full
scaling function for the non-linear conductance.

{\it Fluctuation-dissipation ratio and spin decoherence:
~}
A number of features underlie 
the $V/T$ scaling found here 
for the 
non-linear 
conductance. First, we observe that the 
kernel in the expression for the current
in Eq.~(\ref{current_expression}) is a function
of $\omega/T$ and $V/T$ only:
$[f_L(\omega)-f_R(\omega)]
=\frac{\sinh(V/2T)}{
\cosh(\omega/T)+\cosh(V/2T)
}$.
This is actually a manifestation
of the particular form for the 
fluctuation-dissipation ratio 
of the T-matrix:
\begin{eqnarray}
FDR_{\mathcal{T}} \equiv 
\frac{{\mathcal{T}}^>+{\mathcal{T}}^<}{{\mathcal{T}}^>-{\mathcal{T}}^<} = 
\frac{\sinh(\omega/T)}{
\cosh(\omega/T)+\cosh(V/2T)
} .
\label{fdr_g}
\end{eqnarray}
This 
result
for the fluctuation-dissipation ratio is valid 
for arbitrary bias voltage,
regardless of whether the system is critical,
and {\it for any $N$}.
It appears not to have been recognized before, but it 
can be traced back to the steady state condition\cite{Meir.92}.
In light of Eq.~(\ref{fdr_g}), the $V/T$ scaling
seen in the 
I-V characteristics 
would follow
if the spectral density 
of 
the T-matrix
exhibits simultaneous
$\omega/T$
and $V/T$ scaling.
The latter 
is indeed observed.

\begin{figure}[t!]
\includegraphics[width=0.5\textwidth]{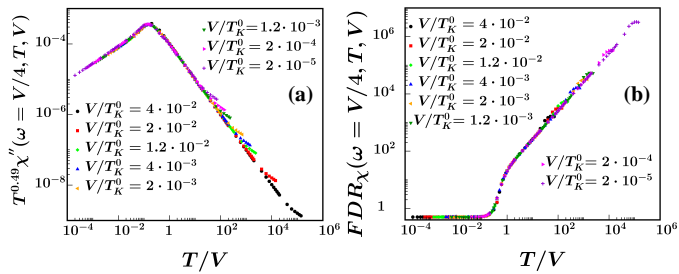}
\caption{Scaling of the local dynamical spin susceptibility.
Shown are the spectral function (a)
and the fluctuation-dissipation ratio (b) defined as 
$FDR_{\chi}=(\chi^>+\chi^<)/(\chi^>-\chi^<)$,
both taken at $\omega=V/4$.}
\label{suszi}
\end{figure}

To explore the origin of such scaling behavior further, and especially 
to elucidate the spin decoherence,
we have studied the local spin
susceptibility. We find that it too displays a $V/T$ and $\omega/T$ scaling. 
Fig.~\ref{suszi}a illustrates the scaling collapse of the spin spectral
density, $\Im\chi^a$. It shows the $V/T$ scaling at a particular 
frequency, $\omega=V/4$; independent $\omega/T$ and $V/T$ scalings 
are obeyed when the $\omega$  and $V$ dependences are separately
analyzed (not shown). 
Fig.~\ref{suszi}b 
shows 
a similar 
scaling collapse for the fluctuation-dissipation ratio of the spin
susceptibility. Note that, unlike for $\mathcal{T}$, the steady
state condition does not fix the fluctuation-dissipation ratio of $\chi$.

Two important points are in order.
First, the violation of the 
fluctuation-dissipation theorem
away from thermal 
equilibrium depends on the 
physical property,
as had already been noted in other settings
({\it e.g.}, Ref.~\cite{Mitra.05}); relatedly,
the notion of an effective temperature apparently
fails here.
Second, the fluctuation-dissipation ratio nonetheless still satisfies
universal scaling.

The $\omega/T$ and $V/T$ scaling occurs only at the quantum
critical point and in the scaling regime of the 
quantum critical LM phase.
As in the equilibrium relaxational regime ($V=0$, $\omega \ll k_BT/\hbar$)
where
an $\omega/T$ scaling implies a linear-in-$T$
spin relaxation rate\cite{Sachdev},
in the non-equilibrium relaxational
regime here ($T=0$, $\omega \ll e V /\hbar $),
the $\omega/V$ scaling implies a linear-in-$V$ dependence of
the decoherence rate,
$\Gamma_V \equiv 
\left [ -i \partial \ln \chi^a (\omega,T=0,V) / \partial \omega
\right ]_{\omega=0}^{-1} = c (e /\hbar) V $.
It is worth stressing
that, in contrast to its counterpart
in the high-voltage ($V \gg k_B T_K^0 /e$) perturbative 
regime\cite{Wingreen.94,Kaminski.00,Paaske.04b},
the decoherence rate here is universal: $c$ is a 
number characterizing the fixed point.

The Kondo effect involving ferromagnetic leads has already been 
observed\cite{Pasupathy.04}, and it was argued that implementing
the tuning to the quantum critical regime is feasible\cite{Kirchner.05a}.
The scaling of the out-of-equilibrium steady-state current 
can accordingly be studied. 
Measuring the local spin response is more challenging 
experimentally. A promising route could be provided by the single-spin ESR
technique\cite{Kroner.08}.

In conclusion, we have identified a system in which quantum
criticality out of equilibrium can be systematically studied.
In this model system, we have shown that the universal scaling is
obeyed by the steady state current, the pertinent spectral
densities, and the associated fluctuation-dissipation ratios.
We have been able to calculate the entire scaling function for each
of these non-equilibrium quantities.
Our theoretical approach can be extended to study the
transient behavior and other non-equilibrium probes of
quantum criticality. 

The work has been supported in part by
NSF Grant No. DMR-0706625, the Robert A. Welch Foundation Grant No. C-1411, 
the W. M. Keck Foundation,
and the Rice Computational Research Cluster
funded by NSF
and a partnership between Rice University, AMD and Cray.

{{\it Note added-} After this paper was posted (arXiv:0805.3717),
another work (C.-H. Chung {\it et al.}, arXiv:0811.1230; 
Phys. Rev. Lett. {\bf 102}, 216803 (2009)) 
has appeared on a nonequilibrium dissipative problem 
with an Ohmic spectrum. Their case has 
a Kosterlitz-Thouless transition, while the sub-Ohmic case we study here
contains 
a genuine quantum critical point. Their results are complementary to ours.

\end{document}